\documentclass{article}
\usepackage{spconf,amsmath,amssymb, graphicx}
\usepackage{enumitem}
\usepackage{bm}
\setlist[itemize]{align=parleft,left=4pt..10pt}

\DeclareUnicodeCharacter{2212}{-}

\title{Multitrack Music Transcription with a Time-Frequency Perceiver}
\name{Wei-Tsung Lu, Ju-Chiang Wang, and Yun-Ning Hung}
\address{SAMI, ByteDance, Mountain View, CA, USA\\
\tt\small \{weitsung.lu, ju-chiang.wang, yunning.hung\}@tiktok.com
}

\begin{document}
\ninept
\maketitle
\begin{abstract}

\end{abstract}
Multitrack music transcription aims to transcribe a music audio input into the musical notes of multiple instruments simultaneously. It is a very challenging task that typically requires a more complex model to achieve satisfactory result. In addition, prior works mostly focus on transcriptions of regular instruments, however, neglecting vocals, which are usually the most important signal source if present in a piece of music. In this paper, we propose a novel deep neural network architecture, \emph{Perceiver TF}, to model the time-frequency representation of audio input for multitrack transcription. Perceiver TF augments the Perceiver architecture by introducing a hierarchical expansion with an additional Transformer layer to model temporal coherence. Accordingly, our model inherits the benefits of Perceiver that posses better scalability, allowing it to well handle transcriptions of many instruments in a single model. In experiments, we train a Perceiver TF to model 12 instrument classes as well as vocal in a multi-task learning manner. Our result demonstrates that the proposed system outperforms the state-of-the-art counterparts (e.g., MT3 and SpecTNT) on various public datasets.


%
\section{Introduction}
\label{sec:intro}

Automatic music transcription (AMT) is a Music Information Retrieval (MIR) task that aims to transcribe a music audio input into a sequence of musical notes where each note contains attributes of onset, pitch, duration, and velocity. The output is typically delivered in the format of MIDI. In a multitrack setting, an AMT system should identify every instrument that is present in the input, and estimate the associated notes accordingly into a channel of the MIDI output. Ideally speaking, using the identified instrument for each corresponding channel, the synthesized audio mixture from the output MIDI should resemble the original input audio in a musically plausible way.

Although recent years have seen significant progress using deep learning techniques \cite{cheuk2021reconvat,gardner2021mt3}, our analysis and review indicate that two major challenges have not yet been addressed effectively: \emph{model scalability} and \emph{instrument discrimination}. 
Multitrack AMT is generally regarded as a very challenging task. The number of commonly used instruments can be up to 100. Among them, musical notes of regular instruments like guitar, violin, and synthesizers are difficult to characterize due to their tremendous variations in timbre, expressivity, and playing techniques. Other than that, vocals, which usually are the most predominant instrument if present, vary their timbre and pitch to convey lyrics and expressions. To handle all instruments simultaneously, it requires better \emph{model scalability}. Our observations on existing multitack AMT systems reveal that they oftentimes result in many false positive notes for popular pitched instruments like piano and guitar. For instance, notes of string ensemble are massively captured by piano. This might be because the system does not provide clear timbre-dependent features or it is not robust to timbral variations across different instruments. We believe this problem can be mitigated if the system can \emph{discriminate} each instrument source from the mixture while making inference.

To address model scalability, we propose \emph{Perceiver TF}, which is an augmented variant of the Perceiver \cite{jaegle2021perceiver}. Perceiver has been well-known for its better scalability in the Transformer family to tackle high-dimensional data input. 
In this work, we adopt spectrogram for audio input, with $T$ and $F$ representing the lengths of the time- and frequency-axes, respectively.   
For multitrack AMT, capability to model the timbre-dependent pitches of multiple instruments is crucial, so more comprehensive operations are needed to capture useful features along the high-resolution frequency axis.
Recently, the SpecTNT architecture \cite{lu2021spectnt} was proposed for this purpose and achieved state-of-the-art performance in vocal melody extraction (a sub-task of AMT). SpecTNT consists of two Transformers in a hierarchical structure, where the lower-level Transformer performs self-attention directly on the spectrum of a frame. However, such design leads to a cubic complexity of attention computation, i.e., $\mathcal{O}(TF^2+T^2)$, limiting its expandability for more complex tasks. To this end, we conceive a non-trivial combination of Perceiver and SpecTNT: expanding Perceiver to be hierarchical. The resulting \emph{Perceiver TF} takes advantage of the cross-attention to extract spectral features into a latent bottleneck for each frame, and adds an additional Transformer for self-attention along the time axis, overall resulting in a quadratic complexity of $\mathcal{O}(TF+T^2)$. Since $F$ is typically large, this complexity reduction is significant, allowing the model to handle more instruments simultaneously.

To address \emph{instrument discrimination}, we adopt the \emph{random-mixing} augmentation technique learned from music source separation (MSS) \cite{uhlich2017improving,song2021catnet}, which aims to separate each instrument stem from the input audio mixture \cite{rafii2018overview}. 
Moreover, we train our AMT model in a multi-task learning fashion, with each sub-task modeling the transcription of an instrument.
This multi-task design along with the random-mixing technique allows more flexibility to train with enormous amount of augmented training samples.
Our strategy differs from previous works that jointly train the AMT task with instrument recognition \cite{hung2019multitask} or MSS \cite{jansson2019joint} to help inform the model of instrument-dependent features. 
To our knowledge, little work has been done using random-mixing technique to improve multitrack AMT.

%

\section{Related work}
\label{sec:format}

Multi-instrument AMT has been explored in several previous works. Wu et al. \cite{wu2020multi} and Hung et al. \cite{hung2019multitask} trained a transcription model with related tasks in a multi-task learning fashion. Tanaka et al. used clustering approaches to separate transcribed instruments \cite{tanaka2020multi}, while Cheuk et al. used unsupervised learning techniques to improve transcription on low-resource datasets \cite{cheuk2021reconvat, cheuk2022jointist}. These prior examples demonstrated that models based on the pianoroll representation are able to capture instrument-dependent onset, pitch, and duration of notes. Different from the pianoroll approach, Gardner et al. \cite{gardner2021mt3} created a new paradigm that proposes a sequence-to-sequence model, called MT3, to tackle multitrack AMT. 
They trained a standard encoder-decoder Transformer to model multitrack MIDI tokens from audio, and demonstrated state-of-the-art performance on several public datasets. 

By contrast, vocal transcription is usually treated as an independent task in the literature, even though it shares the same goal of AMT. Due to the lack of training data, few works focused on transcribing note-level outputs from polyphonic music audio. Recently, Wang et al. released a human annotated dataset including 500 Chinese songs \cite{wang2021preparation}. They provide a CNN based model (EFN) for a baseline of the task. In \cite{kum2022pseudo}, a teacher-student training scheme is proposed to utilize pseudo labels derived from F0 estimations of vocal. Lately, \cite{hsu2021vocano} proposed a vocal transcription system that requires an MSS as front-end. In this work, we propose a unified framework that combines vocal and multi-instrument transcriptions, and it does not rely on pre-trained modules such as an MSS front-end.


\section{Methodology}
\label{sec: methods}

\begin{figure}
\centering
\includegraphics[width=0.93\columnwidth, trim={0.1cm 0.1cm 0.1cm 0.1cm}, clip]{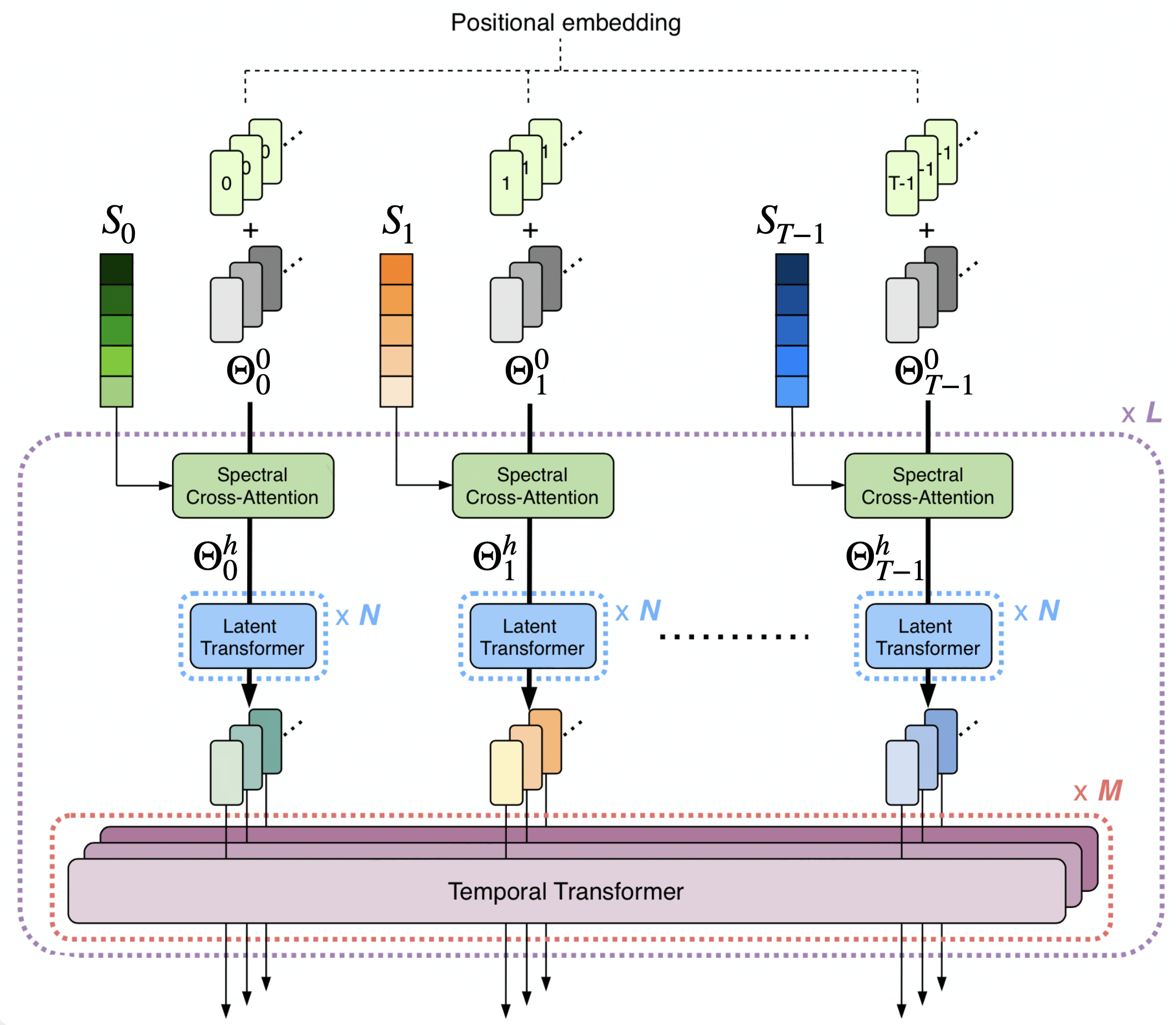}
\caption{The block diagram of the Perceiver TF module. Positional embedding is first added to the latent arrays, denoted as $\Theta^0_t$. The Spectral Cross-Attention module projects the spectral input $S_t$ into $\Theta^h_t$, followed by the Latent Transformer module. The Temporal Transformer processes $\Theta^h_t$ of all time steps to model the temporal coherence. The details are explained in Section \ref{sec:p_tf}.}
\vspace{-0.3cm}
\label{fig:perceiver_tf}
\end{figure}

In this work, we adopt the pianoroll approach instead of the sequence-to-sequence (seq-to-seq) approach for two major reasons. First, it is easier to manipulate the loss computation to learn from partially labeled data. For example, it is non-trivial to train a seq-to-seq model that joints a vocal transcription dataset where the MIDI ground truth of accompaniments is not available. Second, the inference time complexity of seq-to-seq depends on the number of notes (tokens) due to the auto-regressive nature. If the audio input contains many instruments with complex, dense polyphonic notes, the inference will be very slow. Although our proposed model is also a Transformer-oriented architecture, we focus on the encoder part to predict the pianoroll directly.

The following sections explain the proposed model architecture (Sections 3.1 -- 3.3) and the random-mixing augmentation technique (Section 3.4).
Our model consists of three sub-modules: convolutional module, Perceiver TF module, and output module. The input spectrogram is first passed through the convolution module for local feature aggregation. Then, the perceiver TF module, which includes multiple Perceiver TF blocks, extracts the features and outputs the temporal embeddings at each time step. Lastly, the output module projects the temporal embeddings into desired dimensions for pianoroll outputs. 

\subsection{Convolutional Module}

Using convolutional neural network (CNN) as the front-end of Transformer-based models has became a common design choice in speech recognition pipeline \cite{gulati2020conformer}. Previous works have also found that the CNN front-end plays an crucial role in SpecTNT and MIRTransformer for many MIR tasks \cite{lu2021spectnt,won2021transformer,hung2022modeling,wang2022catch}.
Following this practice, we stack multiple residual units \cite{he2016identity} with average pooling to reduce the dimensionality of the frequency axis.
We denote the resulting time-frequency representation as $\mathcal{S} = [S_0, S_1, \dots, S_{T-1}] \in \mathbb{R}^{T \times F \times C}$, where $T$, $F$, and $C$ represent the dimensions of time, frequency, and channel, respectively.

\subsection{Perceiver TF Module}
\label{sec:p_tf}
A conventional Perceiver architecture contains two major components \cite{jaegle2021perceiver}: (i) a cross-attention module that maps the input data and a latent array into a latent array; (ii) a Transformer tower that maps a latent array into a latent array. 
Upon this structure, our design principle to expand Perceiver is twofold. (1) We consider the spectral representation of a time step, $S_t$, is pivotal to carry the pitch and timbral information, so it serves as the input data for the cross-attention module to project the spectral information into a latent array for the time step $t$.
Each latent array is responsible for extracting the local spectral features. 
(2) Having a sequence of latent arrays of different time steps, we need a Transformer to exchange the local spectral information along the time axis to learn their temporal coherence. 

The Perceiver TF architecture is illustrated in Fig. \ref{fig:perceiver_tf}. 
A Perceiver TF block contains three Transformer-style modules: \emph{spectral cross-attention}, \emph{latent Transformer}, and \emph{temporal Transformer}, which are responsible for modeling the spectral, channel-wise, and temporal information, respectively.
Each of them includes the attention mechanism and a position-wise feed-forward network.

The \emph{spectral cross-attention} (SCA) module operates directly on an input spectral representation $S_t$ 
and projects it into the Key ($K$) and Value ($V$) matrices.
Unlike the traditional Transformer, the cross-attention module in Perceiver maps a latent array into the Query ($Q$) matrix and then performs the $QKV$ self-attention accordingly. We follow the Perceiver design to initialize a set of $K$ learnable latent arrays $\Theta^{0} \in \mathbb{R}^{K \times D}$, where $K$ is the index dimension, and $D$ is the channel dimension. Then, we repeat $\Theta^{0}$ for $T$ times and associate each to a time step $t$, which is then denoted as $\Theta_t^0$, such that $\Theta_0^0=\Theta_1^0=\dots \Theta_{T-1}^0$,
meaning that all latent arrays are from the same initialization across the time axis. 
This $\Theta_t^h$ plays an important role of carrying the spectral information from the first Perceiver TF block throughout the entire stack of blocks. The query-key-value ($QKV$) attention of our SCA of the $h$-th iteration can be written as: $f_{\text{SCA}}: \{\Theta_t^h , S_t\} \rightarrow \Theta_t^{(h+1)}$, and this process will be repeated as the Perceiver TF block repeats in order to maintain the connection between $\Theta_t^h$ and the input $S_t$. The design of the cross-attention module is the key that significantly improves the computational scalability of Perceiver. For instance, our SCA results in $\mathcal{O}(FK)$, which is much cheaper than $\mathcal{O}(F^2)$ of the spectral Transformer in SpecTNT \cite{lu2021spectnt}, given that $K$ (dimension of the latent array) is typically small (i.e., $K \ll F$).

The \emph{latent Transformer} module takes place after the SCA module. It contains 
a stack of $N$ Transformers to perform standard self-attention on the latent arrays of $\Theta_t^h$. The resulting complexity $\mathcal{O}(NK^2)$ is efficient as well.
In the context of AMT, this process means the interactions among the onsets, pitches, and instruments are explicitly modeled.
To perform multitrack AMT, we initialize $K$ latent arrays and train each latent array to handle one specific task. Following \cite{hawthorne2017onsets}, for an instrument, we arrange two latent arrays to model the onset and frame-wise (pitch) activations, respectively. This leads to $K=2J$, where $J$ is the number of target instruments. 

The \emph{temporal Transformer} module is placed to enable the communication between any pairs of $\Theta_t^h$ of different time steps. 
To make the \emph{temporal Transformer} understand the time positions of each latent array, we add a trainable positional embedding to each $\Theta_t^0$ during the initialization.
Let $\theta_t^h(k)$, $k$ = 0, $\dots$, $K$--1, denote each latent array in $\Theta_t^h$, we arrange $K$ parallel standard Transformers in which each serves the corresponding input sequence of latent arrays: $[\theta_0^h(k), \theta_1^h(k),\dots,\theta_{T-1}^h(k)]$. The module is repeated $M$ times, yielding a complexity of $\mathcal{O}(MT^2)$.

Finally, we repeat $L$ times the Perceiver TF block to form the overall module. Note that, different from the original Perceiver, the weights of \emph{spectral cross-attention} and \emph{latent Transformer} are not shared across the repeated blocks.

\subsection{Output Module}

We utilize two GRU modules \cite{chung2014empirical} with sigmoid activation function for the onset and frame-wise latent array outputs, respectively. We follow prior work \cite{hawthorne2017onsets} that uses the onset outputs to condition the frame-wise activation.

\subsection{Multi-task Training Loss}

We formulate the loss function for training the proposed model:
\begin{equation}
    \mathcal{L} = \sum^{J-1}_{j=0} (l^j_\text{onset} + l^j_\text{frame})
    \label{eq: loss}
\end{equation}
where $l$ is the binary cross-entropy loss between the ground-truth and prediction, $l^j_\text{onset}$ and $l^j_\text{frame}$ are respectively the onset and frame activation losses for instrument $j$. Note that the losses for all $J$ instruments should be computed, regardless of whether the corresponding instruments are active or not in a training sample. Therefore, a zero output is expected for instruments that are not present in the sample.

\begin{table*}
\renewcommand\thetable{2}
 \begin{center}
 \scalebox{1}{%
 \begin{tabular}{|l||c| c c c c c c c c c c c c |}
 \hline
   Slakh & All & Piano & Bass & Drums & Guitar & Strings & Brass & Organ & Pipe & Reed & S.lead & S.pad & C.perc. \\
  \hline
  MT3$^\dagger$ & .743  & .780 & .906 & .773 & .732 & .551 & .433 & .363 & .282 & .440 & .409 & .234 & .353 \\
  \hline
  \hline
  Ours (No-RM) & .763 & .809 & .921 & .759 & .727 & .699 & .632 & .562 & .578 & .649 & .677 & .358 & .458 \\
  Ours & \textbf{.798} & \textbf{.854} & \textbf{.930} & .\textbf{785} & \textbf{.777} & \textbf{.744} & \textbf{.732} & \textbf{.694} & \textbf{.666} & \textbf{.725} & \textbf{.769} & \textbf{.474} & \textbf{.575}
  \\
  \hline
 \end{tabular}
 }
\end{center}
 \vspace{-0.4cm}
 \caption{The results of different models trained on (Mix) datasets and tested on Slakh2100. MT3$^\dagger$ is our replication, as the instrument-wise results are not reported in \cite{gardner2021mt3}. 
 ``All'' presents the Multi-instrument Onset F1 scores. The following columns show the Onset F1 scores for individual instrument. ``S.lead'', ``S.pad'', and ``C.perc.'' stand for Synth Lead, Synth Pad, and Chromatic Percussion, respectively.}
 \label{tab:inst_wise_perf}
\end{table*}

\begin{table}
\renewcommand\thetable{1}
 \begin{center}
 \scalebox{1}{%
 \begin{tabular}{l|ccc p{1cm}}
 \hline
 Dataset & Slakh & MAESTRO & GuitarSet & MIR-ST500 \\
  \hline
  \hline
  SpecTNT (Single) & - & \textbf{.969} & .907 & .778 \\
  MT3 (Single) & .760 & .960 & .830 & - \\ 
  MT3 (Mix)  & .760 & .950 & .900 & -  \\
  MT3$^\dagger$ (Mix)  & .763 & .958 & .891 & -  \\
  \hline
  Ours (Single)  & .808 & .967 & .903 & .777 \\
  Ours (Mix+Vocal) & \textbf{.819} & .968 & \textbf{.911} & \textbf{.785} \\
  \hline
  EFN & - & - & - & .666 \\
  JDC$_{note}$(L+U)  & - & - & - & .697 \\
  \hline
 \end{tabular}
 }
\end{center}
 \vspace{-0.4cm}
 \caption{The results of Onset F1 scores. MT3$^\dagger$ is our replication. Models with (Mix) or (Mix+Vocal) are trained on the mixture of datasets, while models with (Single) are trained on a single dataset.
 }
 \label{tab:main_perf}
\end{table}

\section{Experiments}
\label{sec: experiments}

\subsection{Datasets}
We use four public datasets for evaluation.
\textbf{Slakh2100} \cite{manilow2019cutting} contains 2100 pieces of multitrack MIDI and the corresponding synthesized audio. The MIDI files are a subset of Lakh dataset \cite{raffel2016learning}, and the audio samples were synthesized by professional-grade software. Instruments were grouped into 12 MIDI classes defined in the Slakh dataset.\footnote{There is no "Sound Effects", "Percussive" and "Ethnic" instruments. We grouped "Strings" and "Ensemble" into one instrument class.} We used the official train/validation/test splits in our experiments.  
\textbf{MAESTROv3} \cite{hawthorne2018enabling} contains about 200 hours of piano solo recordings with the aligned note annotations acquired by the MIDI capturing device on piano. We follow the official train/validation/test splits.
\textbf{GuitarSet} \cite{xi2018guitarset} contains 360 high-quality guitar recordings and their synchronized note annotations. Since there is no official splits for this dataset, we follow the setting in \cite{gardner2021mt3}. The first two progressions of each style are used for training, and the last one is for testing.
\textbf{MIR-ST500} \cite{wang2021preparation} contains 500 Chinese-pop songs with note annotations for the lead vocal melody. We used the official train-test split. Although around 10\% of the training set is missing due to failure links, we ensure the testing set is complete for fair comparison.

\subsection{Data Augmentations}

Annotating data for multitrack AMT is labor intensive. To better exploit the data at hand, we apply two data augmentation techniques during training.
Following previous works \cite{lu2021spectnt, kum2019joint}, pitch-shifting is randomly performed to all the non-percussive instruments during training. 
We introduce the cross-dataset \emph{random-mixing} (RM) technique. Let us first define three types of datasets: 
\begin{itemize} 
    \item \emph{Multi-track}: each sample contains multi-tracks of instrument-wise audio stems with polyphonic notes (e.g., Slakh), and no vocal signals are present.
    \item \emph{Single-track}: each sample contains only a single non-vocal stem with polyphonic notes (e.g., MAESTRO and GuitarSet).
    \item \emph{Vocal-mixture}: each sample is a full mixture of music with monophonic notes only for lead vocal (e.g., MIR-ST500). We employ a MSS tool \cite{kong2021decoupling} to separate each sample into vocal and accompaniment stems. 
\end{itemize}

\noindent Each training sample is excerpted from a random moment of its original song with a duration depending on the model input length (e.g., 6 seconds). Suppose we want to transcribe $J$ classes of instruments, and the corresponding instrument set is denoted as $\mathbf\Omega = \{\omega_j\}_{j=0}^{J-1}$.
Then, we apply three treatments to the three mentioned types of datasets respectively as follows. 

First, for a training sample $s_i$ from a \emph{multi-track} dataset, we denote its instrumentation template as $\bm{\mu}_i \subseteq \mathbf\Omega$, indicating the instruments present in $s_i$. Then, for each instrument $\omega_j$ in $\bm\mu_i$, it has a $p$\% chance to be replaced by a $\omega_j$ in $\bm\mu_u$, where $i \neq u$ (i.e., a different sample).
Second, for a sample $s_i$ from a \emph{single-track} dataset, we randomly pick an existing instrumentation template $\bm\mu_u$ ($i \neq u$) as its background. If the instrument of $s_i$ is present in $\bm\mu_u$, that stem will be removed from $\bm\mu_u$. For instance, if $s_i$ is a piano solo, then we will remove the piano stem from $\bm\mu_u$. From our preliminary experiment, presenting the solo example to model training without mixing it with a background can degrade the performance.
Lastly, for a sample $s_i$ from a \emph{vocal-mixture} dataset, it has a $q$\% chance to replace its background by two methods:
(i) like the \emph{single-track} treatment, we randomly pick an existing $\bm\mu_u$ ($i \neq u$) as its background; or (ii) we randomly pick an accompaniment stem separated from $s_v$, where $i \neq v$. For the second method, since the selected accompaniment stem does not have the ground-truth notes, we mask the instrument outputs and only count the loss for the vocal output (see Eq. \ref{eq: loss}).

\subsection{Implementation Details}

We implemented our system using PyTorch \cite{paszke2019pytorch}. The audio waveform is re-sampled to 16kHz sampling rate. We set the model input length to be 6 seconds. The log-magnitude spectrogram is then computed using 2048 samples of Hann window and a hop size of 320 samples (i.e., 20 ms). The convolutional module contains 3 residual blocks, each of them has 128 channels and is followed by an average pooling layer with a time-frequency filter of (1, 2).

For the Perceiver TF module, we use the following parameters (referring to Fig. \ref{fig:perceiver_tf}): (i) depending on different experiment configurations, initialize $2J$ latent arrays, each uses a dimension of 128; (ii) stack $L=3$ Perceiver TF blocks; (iii) for each Perceiver TF block, use 1 spectral cross-attention layer, $N=2$ latent Transformer layers, and $M=2$ temporal Transformer layers. All the Transformer layers has an hidden size of 128 with 8 heads for the multi-head attention. Finally, the output module is a 2-layer Bi-directional GRU with 128 hidden units.
All of the Transformer module in the Perceiver TF include dropout with a rate of 0.15. 
The output dimension for onset and frame activations are 128 and 129, respectively, where 128 corresponds to the MIDI pitches, and the additional 1 dimension in the frame activation is for the silence. 
We use AdamW \cite{loshchilov2017decoupled} as the learning optimizer. The initial learning rate and weight decay rate are set to $10^{−3}$ and $5 \times 10^{-3}$, respectively.

For final output, we take a threshold of 0.25 for both the onset and frame probability outputs to get the binary representations, so the frame-wise activations can be merged to generate each note in a piano-roll representation. No further post-processing is applied.

For data augmentation, all of the non-percussive instruments of a training example have a 100$\%$ probability to be pith-shift up or down by at most 3 semi-tones. For random-mixing, we use $p=25\%$ and $q=50\%$ for data from \emph{multi-track} and \emph{vocal-mixture} datasets, respectively. To generate an input sample, all the instrument stems in each training example are linearly summed up.



\subsection{Baselines}

Two state-of-the-art models, MT3 \cite{gardner2021mt3} and SpecTNT \cite{lu2021spectnt}, are selected as the baselines. For MT3, we replicated the model following \cite{gardner2021mt3},\footnote{https://github.com/magenta/mt3/blob/main/mt3/colab/\\music\_transcription\_with\_transformers.ipynb} which includes the official model checkpoint and inference pipeline on the test set. 
For SpecTNT, we adopted the configuration used for vocal melody extraction reported in \cite{lu2021spectnt}. In the preliminary experiments, we found it non-trivial to successfully train the original SpecTNT on Slakh2100 under the multi-instrument setting, so we skip this experiment. For vocal transcription, the best results of EFN \cite{wang2021preparation} and JDC$_{note}$(L+U) \cite{kum2022pseudo} are reported.

\subsection{Evaluation Metrics}
We use ``Onset F1'' score, which indicates the correctness of both pitches and onset timestamps, as the evaluation metric for comparison with previous work \cite{gardner2021mt3}. 
To further evaluate the performance of multi-instrument transcription, we report the "Multi-instrument Onset F1" score for the Slakh dataset. The outputs from our replicated MT3 model are grouped into 12 instrument classes based on their program numbers. The Multi-instrument Onset F1 score we used only counts Onset F1, which is similar to the MV2H metric \cite{mcleod2018evaluating}. It could be slightly different from the one used in \cite{gardner2021mt3}, since the ``Drums'' outputs do not contain clear offset information.  
\vspace{-0.5\baselineskip}

\subsection{Result and Discussion}

Table \ref{tab:main_perf} shows the comparison in terms of Onset F1 between the proposed model and baselines.  
The proposed model and SpecTNT which directly model the spectral inputs with the attention mechanism shows higher performance for cases even trained on low resources of a single dataset, such as GuitarSet. On MIR-ST500, the proposed model significantly outperforms the baselines. Although SpecTNT (Single) performs slightly better than our model on MAESTRO, we still consider Perceiver TF to be more advantageous to practical use for its better inference efficiency.


Table \ref{tab:inst_wise_perf} presents the Multi-instrument Onset F1 (instrument-weighted average) and the Onset F1 scores of individual instrument classes on Slakh2100 to reveal instrument-wise performance.
Compared to $\text{MT3}^\dagger$, our model without the random-mixing augmentation (No-RM) performs significantly better on less-common instruments such as ``Pipe'' (the Onset F1 score is upper by over 100\%). Applying random-mixing in training can further boost the performance in all cases, indicating the technique indeed improves the model robustness to discriminate between different instruments.
Finally, we observe that combining multi-instrument and vocal transcriptions can improve the vocal transcription alone, as the combined model is trained with more randomly mixed vocal-accompaniment samples.  

\section{Conclusion}
We have presented Perceiver TF, a novel architecture that adequately addresses the \emph{model scalability} problem for multitrack AMT. To address the \emph{instrument discrimination} issue, we have proposed the random-mixing augmentation technique, which significantly facilitates the data usability across datasets. 
Our system has demonstrated state-of-the-art performance on different public datasets. We believe Perceiver TF is generic and can be applied to other analogous tasks.
\label{sec:majhead}



\bibliographystyle{IEEEbib}
\bibliography{refs}

\end{document}